\documentstyle[prl,aps,multicol,epsf]{revtex}

\begin{document}
 
 \draft
 \title{An observation of spin-valve effects in a semiconductor
 field effect transistor: a novel spintronic device.}
 \author{S. Gardelis, C.G Smith, C.H.W. Barnes, E.H.
 Linfield and D.A. Ritchie}
 
 \address{University of Cambridge,
 Cavendish Laboratory,
 Madingley Road,
 Cambridge CB3 0HE,
 United Kingdom
 }
 
 \date{\today}
 
 \maketitle
 
 \widetext
 \begin{abstract}
 \leftskip 54.8pt
 \rightskip 54.8pt
 
 We present the first spintronic semiconductor field effect transistor.
 The injector and
 collector contacts of this device were made from magnetic permalloy
 thin films with
 different coercive fields so that they could be
 magnetized either parallel or antiparallel to each other in different
 applied magnetic fields.  The conducting medium was a
 two dimensional electron gas (2DEG) formed in an AlSb/InAs quantum well.
 Data from this device suggest that its resistance
 is controlled by two different types of spin-valve effect:
 the first occurring at the ferromagnet-2DEG interfaces;
 and the second occurring in direct propagation between contacts.
 \end{abstract}
 
 
 \begin{multicols}{2}
 \narrowtext

 The idea of electronic devices which exploit both the charge and spin of
 an electron
 for
 their operation has given rise to the new field of `spintronics',
 literally spin-electronics \cite{Lucent,kane98}.  The two-component
 nature of spintronic
 devices is expected to allow a simple implementation of quantum computing
 algorithms as well as producing spin transistors and spin based memory
 devices \cite{Lucent,kane98}.
 However, this new field has yet to have any real impact on the
 semiconductor
 microelectronics industry since no implementation of a spintronic device
 has
 appeared in the form of a semiconductor field effect transistor (FET).
 
 Spin-polarized electron transport from magnetic to non-magnetic metals
 has been the subject of intense investigation since the early 70's
 when Tedrow and Meservey \cite{Tedrow} demonstrated
 the injection of a spin-polarized current from
 ferromagnetic nickel to superconducting aluminium.
 This work was subsequently extended to include spin-dependent transport between
 other materials.  The investigation of
 ferromagnetic-ferromagnetic/paramagnetic materials
 \cite{Julliere,Johnson}
 resulted in the important discovery of the giant magnetoresistance effect
 \cite{Camley,Baibich}.  Work on
 ferromagnetic-semiconductor systems has so far been more limited.
 Alvarado and Renaud \cite{Alvarado}  have demonstrated spin-polarized
 tunneling from a ferromagnet into a semiconductor by
 analyzing luminescence induced by a tunneling current between a nickel tip 
 and a GaAs surface in a scanning tunneling microscope (STM).
 Similar experiments were conducted by Sueoka {\it et al} \cite{Sueoka} 
 and Prins {\it et al} \cite{Prins}.
 
 In this paper, we present results from a spintronic
 semiconductor FET based on the {\it theoretical} ideas of
 Datta and Das \cite{Datta}.
 In their proposed FET, resistance modulation is achieved through
 the spin-valve effect \cite{explain} by varying the degree of spin
 precession which occurs in
 a two dimensional electron gas (2DEG) between identical ferromagnetic
 contacts.
 In our device, resistance modulation is also achieved through
 the spin-valve effect but by having
 ferromagnetic contacts with different coercivities and varying an
 applied magnetic field. We show that the low field
 magnetoresistance of the device results from two types of spin-valve
 effect: a ferromagnet-semiconductor contact resistance; and a direct
 effect between the magnetic contacts.

 The device consisted of a 2DEG formed in a
 15nm wide InAs well between two AlSb barriers.  The top barrier was 15nm
 thick and had a 5nm GaSb cap layer to prevent oxidation of
 AlSb.  Two parallel ferromagnetic permalloy ($Ni_{80}Fe_{20}$)
 contacts (see inset of figure 1a), one 5$\mu$m wide (contact A) and the
 other
 1$\mu$m wide (contact B), were patterned using electron beam lithography.
 They were placed 1$\mu$m apart and stretched across a 25$\mu$m
 wide Hall bar produced by optical
 lithography.   The different aspect ratios of these
 contacts ensured that they had different coercivities with an easy
 axis of
 magnetization along their long axes \cite{Adeyeye}. This
  allowed them to be magnetized either parallel or
 antiparallel to each other in different ranges of external magnetic
field.
 To ensure good ohmic behavior between the contacts A and B
 and the 2DEG, the top GaSb and AlSb layers
 were etched away selectively in the area of the contacts
 using Microposit MF319 developer \cite{Gatzke}. Any oxide on the InAs
 surface, which could act
 as spin scatterer, due to the paramagnetic nature of oxygen, was removed
 by
 dipping the sample in $(NH_{4})_2S$.
 This is known to passivate
 the InAs surface with sulfur and decelerate the oxidation process
 \cite{Oigawa}. Moreover, it
 has been shown to improve the tunneling properties in STM
 studies of InAs \cite{Canali}. It is also expected to
 remove
 any Sb residue which is known to be present after etching AlSb with the
 MF319 developer \cite{Gatzke}.
 Within 5 minutes of passivation, a film of 50nm of permalloy
 was evaporated followed by 20nm of Au in order to protect the
 permalloy from oxidation.
 A network of extended NiCr/Au contacts (shown as 1-8 in figure 1a,
inset),
 patterned by optical lithography, was used to connect
 contacts A and B with external circuitry. A layer of polyamide
 insulated this network from the device surface.
    Non-magnetic
 NiCr/Au ohmic contacts used
 for four-terminal device characterization were
 patterned at each end of the Hall bar.
 For basic non-magnetic characterization an identical
 Hall bar without magnetic contacts was prepared on the same wafer.
 
 A reduction in mobility of the device 2DEG from
 $\mu$=4.9 to 0.09m$^{2}$s$^{-1}$V$^{-1}$ (at 0.3K)
 was observed after removal of the AlSb barrier in the regions of
 contacts A and B and subsequent dipping of the device in $(NH_{4})_2S$.
 A reduction in mobility (from 3.6 to 0.5m$^{2}$s$^{-1}$V$^{-1}$ at 0.3K)
 was observed in a reference sample, having no
 magnetic contacts, after removal of the AlSb barrier above the InAs well
 over
 the whole surface of the Hall bar. We believe that the reduction in
 mobility
 in the device is
 partly due to lateral etching of the AlSb barrier layer located between
 the
 magnetic contacts after dipping in $(NH_{4})_2S$. It is known that
 $(NH_{4})_2S$
 attacks GaSb and AlSb chemically \cite {Andreev}. It is also possible
that
 inhomogeneous band bending at the sulfur-passivated
 surface produces greater charge scattering
 than an oxide surface.
 
 The spin transport properties of the 2DEG are important to the operation of
 the device.  There are two parts to this: a 2DEG in an InAs quantum
 well is
 diamagnetic \cite{Sahu,Smith}; and has strong spin precession 
 \cite{Rashba,Chen}.
 This spin precession results from the Rashba \cite{Rashba}
 term in the spin-orbit interaction.  In transport, the combination of multiple
 elastic scattering from non-magnetic
 impurities
 and spin-precession results in a randomization of spin
 orientation and can give rise to weak
 antilocalization \cite{Chen}.
 This effect was observed in our
 characterization Hall bar (at the center of a weak localisation peak)
enabling
 us to estimate the spin dephasing length, $\ell_{sd}$, and the related
 zero-field spin-splitting
 energy, $\Delta E$, of the device 2DEG.
 These measurements were made in a magnetic field applied
 perpendicular
 to the 2DEG.

 By fitting the weak antilocalization peak as described in ref. \cite{Chen} we
 estimated the spin dephasing time, $\tau_{s}$, to be 9ps.
 For the calculations we used an electron density
 $n = 6 \times 10^{15}$m$^{-2}$, calculated
 from the Shubnikov-de-Haas oscillations, a mobility of
 $\mu$=4.9m$^{2}$s$^{-1}$V$^{-1}$ and the
 effective mass for InAs $m^{\ast}$=0.04$m_o$ ($m_o$ = electron rest mass)
 \cite{Yang}.
 $\ell_{sd}$, was calculated from
 the expression,
 $\ell_{sd}=({\ell\upsilon_{F}\tau_s})^{1/2}$, where $\ell$ is the elastic
 mean free path
 and $\upsilon_F$ is the Fermi velocity in our system, and found to be
 $\ell_{sd}$= 1.8$\mu$m.
 $\Delta E$ at zero magnetic field was calculated using an expression
 for
 $(\tau_s)^{-1}$ given in \cite{Chen},
 $(\tau_s)^{-1}$=$(<\Delta E^2>\tau_e)$/$4\hbar^2$, where $<\Delta E^{2}>$
 is the Fermi-surface
 average of $\Delta E^{2}$, and $\tau_{e}$ is the relaxation time for
 elastic scattering and was found to be
 $<\Delta E^{2}>$= 0.16(meV)$^2$. From the expressions for $\tau_{s}$ and $\ell_{sd}$  we can see that
 $\ell_{sd}={2\hbar\upsilon_{F}}/{\sqrt{<\Delta E^{2}>}}$
 implying that $\ell_{sd}$ is independent of the mobility.
Our device should therefore be expected to have a
 similar $\ell_{sd}$ to the characterization Hall bar since
 they have similar
 carrier concentrations and zero-field spin splitting.

 Weak antilocalization was not observed after sulfur passivation since a
 reduction in mobility causes a
 reduction in the inelastic scattering length $\ell_{\varphi}$ and can
 therefore break
 the condition for observation of weak
 antilocalization ( $\ell_{\varphi}$ comparable or larger than  $\ell_{sd}$ ).
 For our characterization Hall bar we
 estimated $\ell_{\varphi} = 1 \mu$m by fitting the weak localization part
 of the magnetoresistance \cite{Altshuler2}.  From
  the ratio of the mobilities with and without sulfur passivation
 we estimate $\ell_{\varphi} \sim 0.1\mu$m$ << \ell_{sd}$ after sulfur
 passivation \cite{Altshuler}.
 
 In order to determine the magnetic properties of the contacts A and B
 we performed four-terminal magnetoresistance measurements
 at 0.3K using a constant ac current
 of 100$\mu$A. For contact A the current was applied between positions 2
 and 6 (see inset in figure 1a)
 and the voltage drop between contacts 1 and 5 was recorded by lock-in
 amplification techniques. Similarly for contact B the current was applied
 between
 positions 3 and 7 and the voltage drop was recorded between positions 4
 and 8
 of the contact.
 These measurements are shown in figure 1a with the magnetic field
 being applied along the long axis of the contacts A and B.
 The sharp minimum in each curve corresponds to the switching
 of the magnetization  of the contact and therefore occurs at its
 coercive field \cite{Adeyeye}.
 For contact A we measured a coercive field $Hc_{A}$=3.5mT and
 for contact B, $Hc_{B}$=8.5mT.
 
 In order to observe the spintronic properties of the device,
 magnetoresistance measurements were carried out in magnetic fields
 applied parallel to the plane of the 2DEG and along the easy axis of
 the contacts A and B at temperatures ranging from 0.3K to
 10K.
 A constant ac current of 1$\mu$A was applied between positions 1
 and 4 (see inset in figure 1a) of the magnetic contacts
 and the voltage drop between positions 5 and 8 was recorded.
  Figure 1b
 shows these measurements, plotted as the change in the magnetoresistance,
 $\Delta R$:
 \begin{equation}
 \Delta R=R(H)-R(H=0),
 \end{equation}
 from its zero field value $R(H=0)=588\Omega$. $H$ is the applied magnetic
 field.
 At 0.3 K figure 1b shows both an up sweep
 (solid line) and a down sweep (dashed line). The principal features 
 in these sweeps
 are a peak in magnetoresistance between the two
 coercive fields $H_{C_{A}}$ and $H_{C_{B}}$ and a dip on either
 side of this peak.  The dip on  the
 low-field side is deeper than the one on the high-field side.
 This structure is repeated
 symmetrically on opposite
 sides of zero field for the up and down sweeps.
 
 By comparing four-terminal resistance measurements made between
 the contacts A and B with those made between the non-magnetic contacts
  the interface conductance, G, was found to be
 10mS.
 Furthermore, the spin conductance of the 2DEG, $g_s$, defined
 as the conductance of a length of the bulk material equal to $\ell_{sd}$,
 \cite{Johnson2}, was found to be 2mS.
 Therefore, since G and $g_s$ are comparable we expect a contribution
 from both the interface and the 2DEG in the device
 magnetoresistance.
 The magnetoresistance ( $\Delta R$ )
 of our device
 will therefore have the following contributions:
 \begin{equation}
 \Delta R=\Delta R_{A} + \Delta R_{B} + \Delta Rc_{A} +\Delta Rc_{B} +
 \Delta R_s\
 \end{equation}
 where $\Delta R_{A}$ and $\Delta R_{B}$ are the magnetoresistance changes
 of
 contacts A and B respectively, $\Delta Rc_{A}$ and
 $\Delta Rc_{B}$ are those of the interface between the 2DEG
 and contacts A and B respectively, and $\Delta R_s$
 is the resistance change due to electrons propagating from one
 ferromagnetic
 contact to the other without spin scattering.
 
 As can be seen by comparing figures 1a and 1b
 the contributions $\Delta R_A$ and $\Delta R_B$
 ($\simeq$ 2m$\Omega$) are 500 times smaller than the magnetoresistance
 changes
 in $\Delta R$ ($\simeq$ 1$\Omega$). The results in figure 1b cannot
 therefore be attributed to changes in the magnetoresistances of the
 contacts themselves.
  The part of the interface resistance $\Delta Rc_A$+$\Delta Rc_B$ 
 which results from the 
 spin-valve effect \cite{explain}, and therefore has a dependence on 
 applied field, will have the schematic form shown in 
 figure 2a.  
  Its shape derives from both the spin properties of the 
 2DEG and the difference in coercive fields of the contacts A and B.  
 It is a
 maximum when the magnetizations of contacts A and B are
 parallel to each other and antiparallel
 to the spin orientation of the 2DEG.   
 It has a minimum value when the
 magnetizations in A and B are both parallel to the spin orientation in
 the 2DEG and an intermediate value when the contact
 magnetizations are antiparallel.
  The part of the resistance contribution from direct propagation between contacts 
 A and B, $\Delta R_s$, which results from the spin-valve effect 
 \cite{explain} will have the form shown schematically in figure 2b.
  This resistance is a minimum
 when both ferromagnetic contacts are magnetized parallel to each other
 and maximum between the two coercive fields where the magnetization of
 the two contacts is antiparallel to each other.
  The broken
 lines in figure 2 represent a more realistic picture of the
 magnetoresistance changes resulting from the two spin-valve effects.
 They represent an average over the local switching of different
 magnetic
 domains
 in the ferromagnetic contacts.
 A schematic representation of the sum of the two spin-valve
  contributions to $\Delta R$
 is shown as a grey line in figure 1b, taking the coercive fields from
 the contact magnetoresistances in figure 1a. This line has the same
 shape as the experiment and appears in the correct place for
 both up and down field sweeps.  The depth and width of the high-field
 magnetoresistance dip
 depend upon the extent to which the shape of the up peak in figure 2b exactly
 compensates the dip down between the coercive fields in figure 2a.
 If these are identical there will be no high-field dip.
 
 The small amplitude of the device resistance modulation
 $\Delta R$/R(H=0)$\approx$0.2$\%$ shown in figure 1b
 is consistent with
 the above picture.  Electrons contributing to the direct spin-valve
 effect shown in figure 2b have to take fairly direct paths
 between contacts A and B.
 Those which take paths involving multiple scattering pick up
 random angles of spin orientation and therefore on average will
 cancel with each other and not contribute to the effect.
 The temperature dependence
 of the magnetoresistance is also consistent with our picture. 
 The peak between the two
 coercive fields
 decreases in
 amplitude with increasing temperature and almost disappears at 10K
 (see figure 1b).  At this temperature  $k_B T$ ( = 0.8meV ) is
 greater than the zero-field spin splitting
 and therefore thermal activation has sufficient energy to
 destroy both spin-valve effects shown in figure 2b.
 
 Alternative mechanisms which could produce the magnetoresistance
 oscillations observed would have to be capable of: producing the
 symmetry we see in up and down field sweeps (solid and dashed lines in
 figure 1b); producing features of an appropriate shape which
 align with the contact coercive fields; and persist up to temperatures
 of 10K in a 2DEG with a zero field resistance of 588$\Omega$ and an
 inelastic scattering length one tenth the length of the device.
 Such
 fluctuations are unknown in the literature.  Universal conductance
 fluctuations (UCFs) could occur in a device of such low 
 resistance. However, their period in magnetic field
 ($H_{C_{B}} - H_{C_{A}}$ = 5mT) is consistent
 with a phase coherent area of $\sim 1 (\mu m)^{2}$ which is two orders of
 magnitude larger than that estimated from the inelastic scattering length
 of the device ( $\ell_{\varphi}\sim 0.1\mu$m). 
 In addition UCFs are not seen in magnetic fields applied parallel 
 to a 2DEG \cite{Kaplan,Taylor}.  Also, since the field was applied 
 along the easy axis of the contacts A and B there should be no 
  stray fields with a significant component perpendicular to the 2DEG. 
   The most likely origin of the small amplitude random modulations appearing
 in the data and the differences
 in shape between up and down magnetic field sweeps is the complex
 pattern of domain formation in the contacts A and B and their pattern of
 switching as a function of external field.

 In conclusion, we have provided evidence that we observed experimentally two kinds of
 spin-valve effect in a spintronic FET.  The first effect results from
the ferromagnet-2DEG interface
 resistance and the second effect results from spins propagating from one
 ferromagnetic contact to the
other. The combination of these effects produces a resistance maximum
 between the coercive fields of the two
 contacts and dips in resistance on either side.
 Both effects are suppressed with increasing temperature as the thermal
 smearing
 becomes comparable to the zero field spin splitting.
 
 We thank J.A.C. Bland, M. Pepper and C. J. B Ford for invaluable
 discussions.  This work was funded under EPSRC grant GR/K89344,
 and the Paul Instrument Fund. CHWB,EHL and DAR acknowledge
 support from the EPSRC the Isaac Newton Trust, and Toshiba Cambridge
 Research Center. The corresponding author: S. Gardelis, email: sg234@cus.cam.ac.uk

 \centerline{\bf Figure Captions}
 
 Fig.1 (a)  Change in resistance of contacts A and B
 with external magnetic field H, averaged
 over 4 up sweeps.
 $H_{C_A}$ and $H_{C_B}$ coercive fields of A and B.
 Inset: schematic of device: black pads - magnetic contacts A,B;
 dark grey - NiCr/Au ohmic contacts; light grey - Hall bar
 mesa.
 (b) Change in device resistance at 0.3K
 averaged over 9 sweeps (up - solid, down - dashed lines) and 4.2K
 and 10K averaged over 2 up sweeps.
  Grey line:
 schematic showing the expected sum of the two spin-valve effects.
 The arrows indicate the direction of the field
 sweep. All traces are offset for clarity.
 
 Fig.2 (a) Schematic of the interface spin-valve effect:
 $\Delta
 Rc_A$+$\Delta
 Rc_B$. Arrows indicate magnetization direction in A and B and
 2DEG S. H is the external field which is being swept up from negative
 value.
 (b) Schematic of direct spin-valve effect: $\Delta R_s$.  Dashed
 lines in (a), (b) indicate 
 averaging over the local switching of different magnetic domains
 in A and B.
 
 \end{multicols}
 

\begin{references}
 
 \centerline{\bf References}
 
 \bibitem{Lucent} The term `spintronics' was apparently
 coined by Lucent Technologies as: `electronic devices
 in which the direction an electron spin is pointing is just as
 important as its charge'. http://public1.lucent.com/press/ 
 0798/980731.bla.html
 
 \bibitem{kane98}B.E.Kane, Nature, {\bf 393}, May 14 133 (1998).
 
 \bibitem{Tedrow} P.M. Tedrow and R. Meservey, Phys. Rev. Lett. {\bf 26},
 192 (1971).
 
 \bibitem{Julliere} M. Julliere, Phys. Rev. Lett. {\bf 54A}, 225 (1975).
 Lett. {\bf 66}, 1926 (1991).
 
 \bibitem{Johnson} M. Johnson and R.H. Silsbee, Phys. Rev. Lett. {\bf 55},
 1790 (1985).
 
 \bibitem{Camley} R.E. Camley and J. Barnas, Phys. Rev. Lett. {\bf 63},
664
 (1989).
 
 \bibitem{Baibich} M.N. Baibich {\em et al.\/}, Phys. Rev. Lett. {\bf 61},
 2472 (1988).
 
 \bibitem{Alvarado} S.F. Alvarado and P. Renaud, Phys. Rev. Lett. {\bf
68},
 1387 (1992).
 
 \bibitem{Sueoka} K. Sueoka, K. Mukasa, and K. Hayakawa, Jpn. J. Appl.
 Phys. {\bf 32}, 2989 (1993).
 
 \bibitem{Prins} M.W.J. Prins, D.L. Abraham, and H. van Kempen, Surf. Sci.
 {\bf 287/288}, 750 (1993).
 
 \bibitem{Datta} S. Datta and B. Das, Appl. Phys. Lett. {\bf 56}, 665
 (1990).
 
 \bibitem{explain} The spin-valve effect \cite{Datta} is an increase
 in resistance experienced by a current of spin polarized electrons
 passing into a material of the opposite magnetization or a reduction
 of resistance in passing into a material of the same magnetization.
 
 \bibitem{Adeyeye} A.O. Adeyeye et al, J. App. Phys. {\bf 79}, 6120
(1996).
 
 \bibitem{Gatzke} C. Gatzke et al, Semic. Sci. Techn. {\bf 13}, 399
(1998).
 
 \bibitem{Oigawa} H. Oigawa et al, Japn. J. Appl. Phys. Part2-Letters {\bf
 30}, L322 (1991).
 
 \bibitem{Canali} L. Canali et al, Proceedings of the 9th International
 conference on scanning
 tunneling microscopy/spectroscopy and related topics (STM'97). Submitted
 to Appl. Phys.
 A- Material Science and Processing.

 \bibitem{Andreev} I.A. Andreev et al, Sov. Phys. Semicond. {\bf 31}, 556
(1997).

\bibitem{Sahu} T. Sahu, Phys. Rev. B {\bf 43}, 2415 (1991).
 
 \bibitem{Smith} T.P. Smith III and F.F. Fang, Phys. Rev. B {\bf 35}, 7729
 (1987).

 \bibitem{Rashba} E.I. Rashba, Sov. Phys. Solid State {\bf 2}, 1109
(1960).

 \bibitem{Chen} G.L. Chen et al, Phys. Rev. B {\bf 47}, 4084 (1993).
 
 \bibitem{Yang} J. Yang et al, Phys. Rev. B {\bf 47}, 6807 (1993).
 
 \bibitem{Altshuler2} B.L. Altshuler, D.E. Khmelnitskii, A.I. Larkin, and
 P.A. Lee,
 Phys. Rev. B {\bf 22}, 5142 (1980).
 
 \bibitem{Altshuler} B.L. Altshuler, A.G. Aronov, and D.E. Khmelnitskii,
J.
 Phys. C {\bf 15}, 7367 (1982).
 
\bibitem{Johnson2} M. Johnson and R.H. Silsbee, Phys. Rev. B {\bf 37},
 5326 (1988).
 
\bibitem{Kaplan} S.B. Kaplan and A. Hartstein, Phys. Rev. Lett. {\bf 
   56}, 2403 (1986).
  
\bibitem{Taylor} R.P. Taylor et al, J. Phys.:Condens. Matter {\bf 1},
  10413 (1989).


 
 \end{references}
 
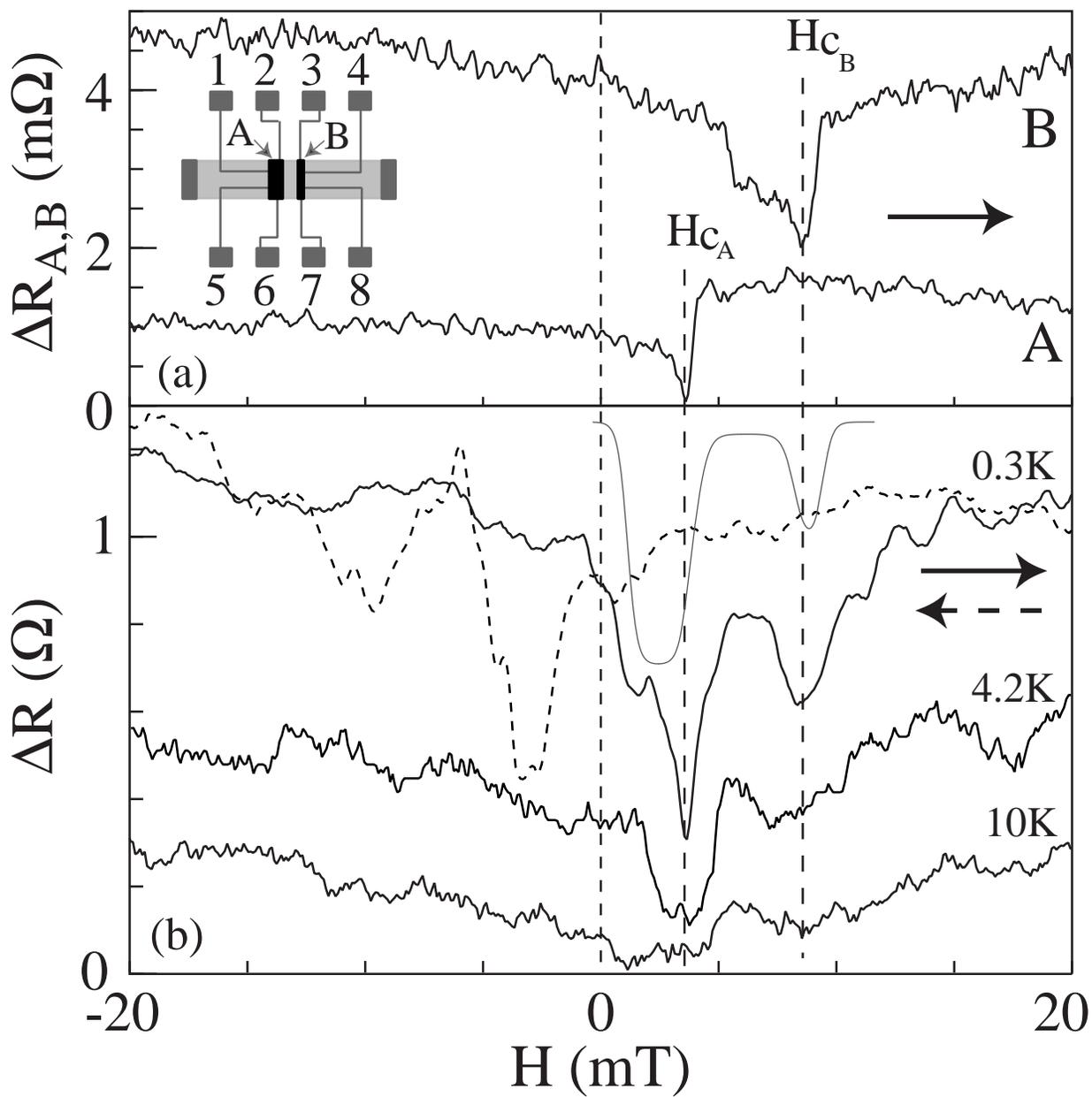
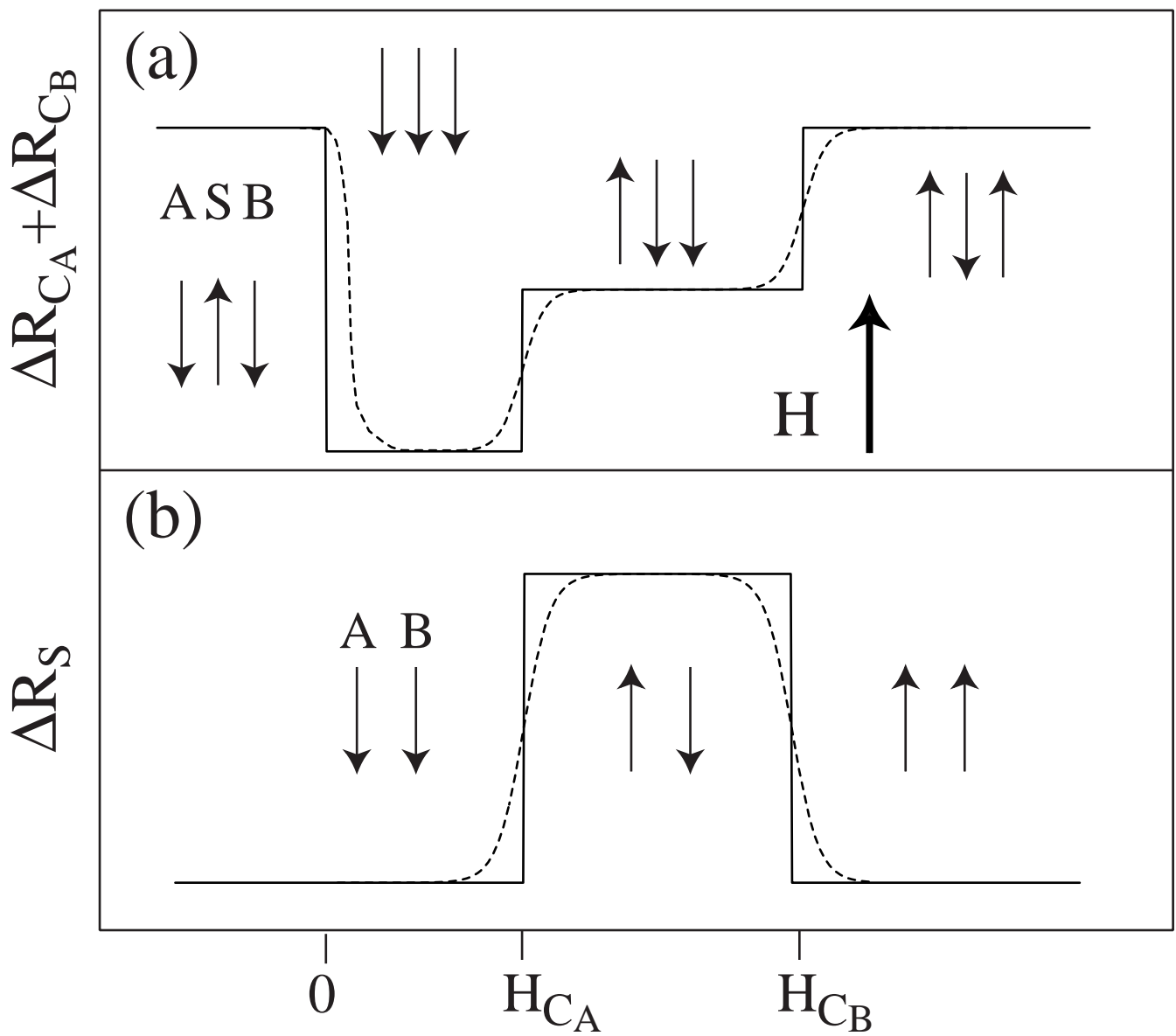
\end{document}